\documentclass[amsmath,amssymb,aps,eqsecnum,showkeys,dvips]{revtex4}
\usepackage[dvips]{graphicx} 

\begin{document}

\title{Euroattractor: a brief introduction to Iterated Function Systems}
 
\author{Karol {\.Z}yczkowski}\email{karol@cft.edu.pl}
\affiliation{Centrum Fizyki Teoretycznej PAN, Al. Lotnik{\'o}w 32/46,
02-668 Warszawa, Poland}
\author{Artur {\L}ozi{\'n}ski}\email{lozinski@if.uj.edu.pl}
\affiliation{Instytut Fizyki im. Smoluchowskiego, Uniwersytet
Jagiello{\'n}ski, ul. Reymonta 4, 30-059 Krak{\'o}w, Poland}

\begin{abstract} In this work we propose a
definition of an {\sl Euroattractor}: an attracting  invariant measure
of a certain iterated functions system (IFS).  An IFS is
defined by specifying a set of functions, defined in subsets of
${\mathbb R}^N$ or in a classical phase space, which act randomly on the
initial point, so it may be considered as a generalization of the notion
of classical dynamical system.  If the functions are sufficiently
contracting, there exists an invariant measure of the system, often
concentrated on a fractal set. We investigate invariant measures of
a certain class of generalized or weakly contracting IFS's.

\end{abstract}


\keywords{stochastic dynamics, iterated function systems, fractals}
\maketitle

\section{Introduction}

Any function $f$ which maps a phase space $\Omega$ into itself may be
considered as a classical dynamical system.  An arbitrary point $x \in
\Omega$ determines the trajectory uniquely, $x_n=f^n(x_0)$, where the
natural index $n$ plays the role of time.  Dynamical system defined by
$f$ is deterministic.  There exist different ways to introduce randomness
into the system.  One possibility relays on taking into account an
additive noise, $x_{n+1}=f(x_n)+\xi_n$, where $\xi_n$ are uncorrelated
random variables drawn according to a prescribed probability
distribution $P(\xi)$.  The deterministic case is obtained if $P(\xi)$
tends to the Dirac delta distribution, $P(\xi)=\delta(\xi)$.

In this work we analyze another class of stochastic systems, in which
the choice of a system is random at each step.  Such systems are called
{\sl Iterated Function System} (IFS), and may be considered as a
generalization of the notion of a classical dynamical system. 

An IFS is defined \cite{Ba} by a set of $k$ functions $f_i:\Omega\to
\Omega$, which represent dynamical systems in the classical phase space
$\Omega$.  The functions $f_i$ act randomly with given probabilities
$p_i$. They characterize the likelihood of choosing a
particular map at each step of the time  evolution of the system.  In
general, the probabilities may be place dependent.

Analyzing a given iterated function system, one may ask, how an initial
point $x_0\in \Omega$ is transformed by the random process.  One may
also consider a more general question, how does a probability measure
$\mu$ on $\Omega$ change with respect to the Frobenius-Perron operator
$M$ associated with the IFS.  If the functions $f_i$ are strongly
contracting, then there exists unique invariant measure $\mu_{*}$ of
$M$ -- see  for instance \cite{K81,E87,Ba,BDEG88} and references
therein.

Interestingly, for a large class of IFS's the invariant measure $\mu_*$
is characterized  by a fractal structure.  If the invariant measure is
attracting, such  IFS's may be used to generate fractal sets in the
space $\Omega$. In particular, there exist iterated function systems
leading to popular fractal sets: Cantor set, Sierpinski carpet or
Sierpinski gasket \cite{Ba}.  In fact by considering a slightly
generalized class of IFS's it is possible to design a random system, the
invariant measure of which possesses any desired properties.  In this
work we shall visualize the statement, by presenting an Euroattractor -
a system, the attractor of which has the shape resembling the contours
of Europe.

This paper is organized as follows.  In the next section we recall the
definition and basic properties of the standard IFS.  Section III is
devoted to a weakly contracting IFS, which exhibit nontrivial invariant
measures, although they do not fulfill the standard contractivity
assumptions.  In section IV we define a generalized class of the systems
and present the definition of an Euroattractor.

\section{Iterated Function Systems}

An iterated function system (IFS) is defined by a set of $k$ functions
(dynamical systems) $f_i$, $i=1,...,k$, defined on a compact space
$\Omega$ and the set of $k$ positive probabilities $p_i$, which sum to
unity.  In the IFS of the first kind the probabilities are constant,
while in the more general, IFS of the second kind \cite{BDEG88,SKZ00},
they are functions of the position, $p_i=p_i(x)$. For any $x\in \Omega$
the condition  $\sum_{i=1}^k p_i(x)=1$ has to be fulfilled.  In general,
IFS $\cal F$ is defined by a set ${\cal F}=\{\Omega, f_i:\Omega\to
\Omega, p_i:\Omega\to[0,1] \}$ with $i=1,\dots,k$.

Any IFS generates a Frobenius--Perron operator $M$, which describes the
time evolution of the probability densities on $\Omega$. In the simplest
case, in which the space $\Omega$ is an interval in $\mathbb R$ and  the
maps $f_i$ are invertible,  any initial probability density $\rho(x)$ is
mapped into $\rho'(x)$ by the Frobenius--Perron operator \cite{LM94}
\begin{equation}
\rho'(x)=M[\rho(x)] = \sum_{i=1}^k p_i \bigl(f_i^{-
1}(x)\bigr)\rho\bigl(f_i^{-1}(x)\bigr) \left| \frac{d f_i^{-1}(x)}{dx}
\right|.  \label{FP1} \end{equation}
If $\rho(x)$ is a non-singular
probability density, so is $\rho'(x)$. However, in the limit $n\to
\infty$ the image $M^n[\rho(x)]$ needs not to be a smooth probability
density, so it is advantageous to work in the larger space of
probability measures on $\Omega$. This space includes also also singular
measures like the Dirac delta distributions and their convex
combinations. 

If an IFS is sufficiently contracting, i.e. 

a) the functions $f_i$ satisfy the Lipschitz condition,
	 $d(f_i(x),f_i(y)) \le L_i d(x,y)$) for any $x, y \in \Omega$.   Here
	 $d(x,y)$ denotes the distance between both points and the Lipschitz
	 constants $L_i$ are assumed to be smaller than unity, 

b) the probabilities $p_i(x)$ are continuous functions, strictly
	  positive for any $x\in\Omega$,
the IFS is called {\sl hyperbolic} and there exists a unique {\sl
invariant probability measure} $\mu_*$ satisfying the equation $M\mu_* =
\mu_*$ \cite{LM94}.

Interestingly, the conditions a) and b) are not necessary to assure the
existence of a unique invariant probability measure.  Some other less
restrictive sufficient assumptions which regard also IFS of the second
kind, with place dependent probabilities,   were analyzed in
\cite{K81,E87,BE88,BDEG88,LY94,JL95,E96,Sz00}.

The invariant measures of an IFS display often fractal properties.  For
instance, consider an IFS of the first kind defined on an interval and
consisting of $k=2$  affine transformations, \begin{equation} {\cal
F}_1=\{\Omega=[0,1], \ f_1(x)=x/3, \ f_2(x)=x/3+2/3; \quad
p_1=p_2=\frac{1}{2} \}.  \label{IFS1} \end{equation} Since both
functions are continuous contractions with Lipschitz constants
$L_1=L_2=1/3<1$, and the probabilities $p_i$ are constant and nonzero
this IFS is hyperbolic. Thus there exists a unique, attracting invariant
measure $\mu_*$. It is not difficult to show \cite{Ba}  that $\mu_*$  is
concentrated uniformly on the Cantor set of the fractal dimension $D=\ln
2/\ln 3$. 

Although $\mu_*$ is singular, it is possible to give a general
prescription, how to compute integrals with respect to this invariant
measure.  This is due to the fact that the measure $\mu_*$ is {\sl
attractive}, i.e., $M^n{\mu}$ converges weakly to $\mu_*$ if $n \to
\infty$.  Therefore, in order to obtain the exact value of an integral
$\int_{\Omega} h(x)~d{\mu_*}$ for any continuous function $h(x)$ it is
sufficient to find the limit of the sequence $h_n:=\int_{\Omega} h(x)~d(
M^n{\mu})$ for an arbitrary initial measure $\mu$.  This method of
computing integrals over the invariant measure $\mu_*$ is purely  {\sl
deterministic} \cite{Ba,E96,Sl97,SKZ00}.  Alternatively, one may employ a
{\sl random iterated algorithm} by generating by the IFS a random
sequence $x_j\in \Omega$, $j=0,\dots,n-1$ which originates from an
arbitrary initial point $x_0$.  Due to the the ergodic theorem for IFS's
\cite{K81,E87,IG90} the mean value $\frac{1}{n}[\sum_{j=0}^{n-1}
h(x_j)]$ converges in the limit $n \to \infty$ to the desired integral,
$\int_{\Omega} h(x)~d{\mu_*}$, with probability one.

To show an example of an IFS of the second kind let us change
probabilities in ${\cal F}_1$ and define 

\begin{equation} {\cal F}_2=\{\Omega=[0,1], \ f_1(x)=x/3, \
f_2(x)=x/3+2/3; \quad p_1(x)=x, \ p_2(x)=1-x \}.  \label{IFS2}
\end{equation} The probability $p_2$ vanish at the point $x=1$, so this
IFS is not hyperbolic. In spite of this fact there exists the unique
invariant measure $\mu_*$ concentrated in a non-uniform way on the
Cantor set \cite{SKZ00}. In this case $\mu_*$ displays multifractal
properties, since the generalized dimension of the measure depends on
the R{\'e}nyi parameter \cite{BS93}.

\section{Weakly contracting IFS} 

Contraction conditions a) and b) imply existence of a unique invariant
measure for the IFS.  However, these conditions are too strong, and
there exist IFSs with an attracting invariant measure which do not obey
them.  If the space $\Omega \subset {\mathbb R}^m$ and the functions
$f_i$ are linear, the Lipschitz condition a) is equivalent to the
statement that the moduli of all elements of the Jacobi matrices
$J_i:=\partial f_i(x_1,...,x_m)/\partial (x_1,...,x_m)$ are strictly
smaller then one.

This is not the case for the following IFS defined on the square
\begin{equation} {\cal F}_3  = \left\{ \begin{array}{r c l}
\Omega=[0,1]\times [0,1], \   k & = &2; \quad  p_1=p_2  =1/2 \\ f_1(x,y)
& =& ( x  y, ~ y) \\ f_2(x,y) & =& ( (x-1)y+1, ~  y+\delta) \end{array}
\right\} , \label{IFS3} \end{equation} since the Jacobi matrices
\begin{equation} J_1=\left( \begin{array} [c]{cc} y & x\\ 0 & 1
\end{array} \right) {\rm \quad and \quad} J_2=\left( \begin{array}
[c]{cc} y & x-1\\ 0 & 1 \end{array} \right) \label{Jacobi}
\end{equation} contain entries equal to unity.  This reflects the fact
that both functions $f_1$ and $f_2$ are not contracting in the
$y$--direction.  Numerical experiments suggest that in spite of this
fact for irrational values of $\delta \ll 1$, the associated
Frobenius--Perron operator $M$ possess a unique invariant measure
$\mu_*$.  Such an IFS, not fulfilling the condition a) or b) will be
called {\sl weakly contracting}.
 
\begin{figure}[hbt] \begin{center}
\includegraphics[width=0.6\textwidth,height=0.5\textwidth]{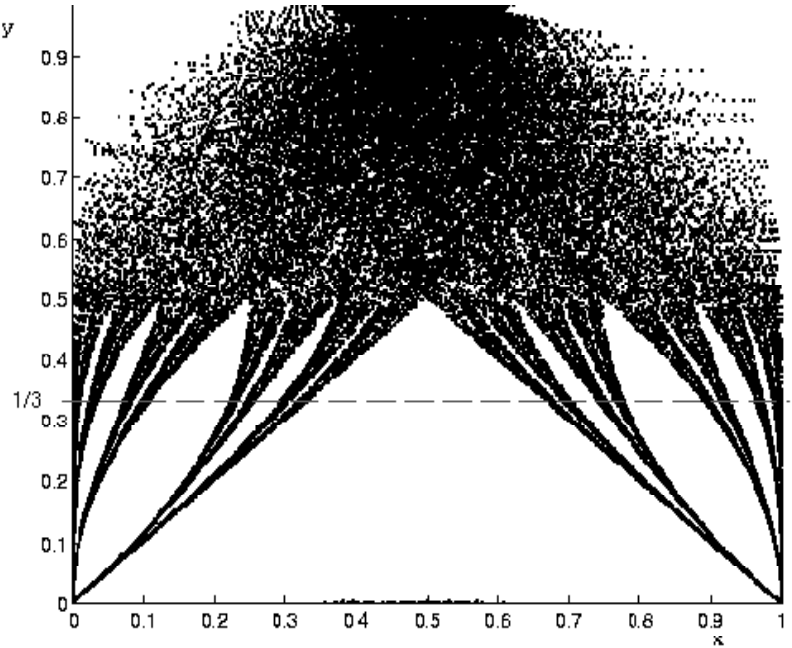}
\end{center} \caption{Invariant density of the weakly contracting IFS
${\cal F}_3$ (\ref{IFS3}) obtained for $\delta=\sqrt{5}/1000$.  The
cross-section of its support at $y=1/3$, denoted by a dashed line,
forms the Cantor set.}
\label{fig1} \end{figure} 

 Each random trajectory of the IFS ${\cal F}_3$ exhibits a slow
diffusion in the $y$--direction, the speed of which is governed by the
rotation parameter $\delta$. Due to periodic boundary conditions,
$(x,y)=(x|_{{\rm mod} 1}, y|_{{\rm mod} 1})$, every trajectory is
confined to the square $\Omega$.  If the parameter $\delta$ is chosen to
be irrational, there are no periodic trajectories.  If $\delta$ is
sufficiently small than the contraction in the $x$--direction will dominate
the diffusion in $y$, so the trajectory will be sticked to the support
of the invariant measure $\mu_*$.  Figure 1 shows a typical trajectory
consisting of $6000$ points generated by this IFS.  To present the
asymptotic properties of the invariant measure the first $100$ points
were omitted.

Note that if the parameter $\delta$ is set to zero, the variable $y$
would be fixed to the initial value $y_0$.  Such an IFS would have an
invariant measure concentrated on the interval $y=y_0$.  Analyzing the
contraction constant we see that for $y_0=1/3$ its support is equal to
the Cantor set generated by the Cantor IFS (\ref{IFS1}). In general, for
any $y_0<1/2$ its dimension would be
\begin{equation} D(y_0) = - \frac
{\ln 2}{\ln y_0} \; . \label{dim1} \end{equation} We conjecture that this
formula gives also the fractal dimension of the horizontal cross-section
at $y=y_0$ of the invariant set of the IFS (\ref{IFS3}) in the limit
$\delta\to 0$. 

Another example of an invariant measure of a nonlinear, weakly
contracting IFS is presented in Fig.2.  The system is defined by  
  
\begin{equation} {\cal F}_4  = \left\{ \begin{array}{r c l}
\Omega=[0,1]\times [0,1], \   k & = &2; \quad  p_1=p_2  =1/2 \\ f_1(x,y)
& =& (\frac{1}{2} x \sin^2(\pi  y), ~ y) \\ f_2(x,y) & =& (\frac{1}{2}
(x-1)\sin^2(\pi  y)+1, ~  y+\delta) \end{array} \right\} .  \label{IFS4}
\end{equation}

As in the previous example each random realization of the IFS generates
a trajectory confined in the torus, and for small values of the shift
parameter $\delta$ the contraction in the horizontal direction dominates
the vertical diffusion. For $\delta=0$ the IFS reduces to
one-dimensional, and the fractal dimension of its invariant set reads
\begin{equation} D (y_0) =  \frac {\ln 2}{\ln2 -2\ln \sin(\pi/y_0)}.
\label{dim2} \end{equation} This number is smaller than one for all
initial values apart of $y_0=1/2$.  Numerical results performed for
$\delta \ll 1$ suggest that the same formula describes the dimension of
the cross-section at $y=y_0$ of the invariant set of the IFS
(\ref{IFS4}) in the limit $\delta \rightarrow 0$. In other words, the
support of the invariant measure $\mu_*$ of the IFS ${\cal F}_4$ may be
considered as a collection of Cantor sets, with a continuously varying
dimension $D(y_0)$, placed horizontally, one parallel to another.

\begin{figure}[hbt] \begin{center}
\includegraphics[width=0.6\textwidth,height=0.5\textwidth]{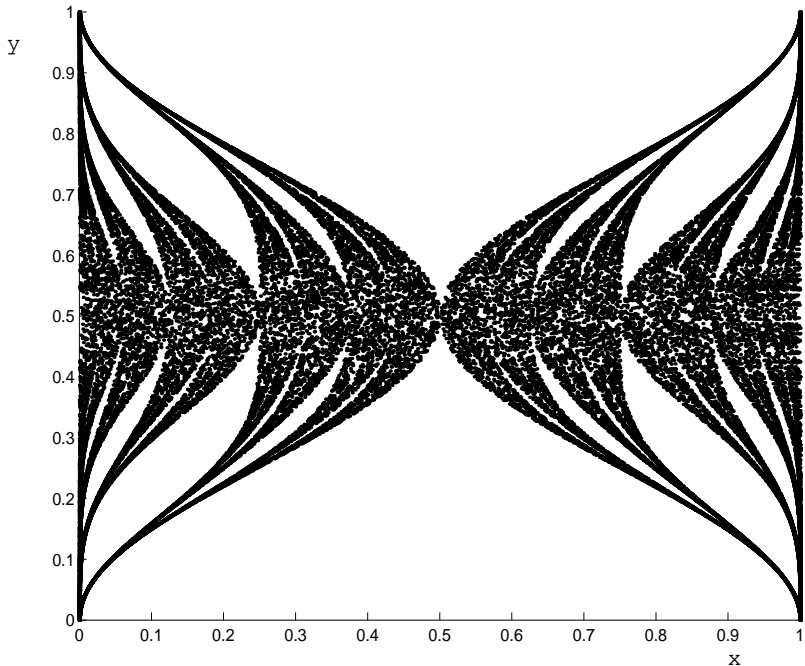}
\end{center} \caption{Invariant density of the weakly contracting IFS
(\ref{IFS4}) obtained for $\delta=\sqrt{5}/1000$.  } \label{fig2}
\end{figure}

\section{Repellers and generalized IFS} 

In the original definition of the IFS, presented in section II, all
functions $f_i$ map the space $\Omega$ into itself. One may, however,
relax this condition, and consider generalized IFS constructed of
functions which map some fragments of $\Omega$ out of this space.  Such
functions are called {\sl repellers}, since they do not conserve the
probability, and during the time evolution the probability flux leaks
out of the system.

A simple $1$-D repeller is obtained from the standard logistic map,
$f(x)=r x (1-x)$, if the control parameter $r$ exceeds the critical
value, $r_c=4$ \cite{Ra89}.  In such systems a randomly selected initial
value leads to a trajectory, which escapes the system
with probability one.  The complementary set of initial points which
generate infinite trajectories is of measure zero and often forms a
fractal set. The same property is characteristic  to a large class of
chaotic 1-D maps with a gap - a subset of $\Omega$ mapped out of
$\Omega$.  An analysis  of the fractal dimension of the invariant set
for the tent map with a gap is presented in \cite{ZB99}.

In this chapter we will consider an IFS defined on the square,
$\Omega=[0,1] \times [0,1]$, consisting of $k$ repelling functions.
Assume that each function $f_i$ restricted to a given set $X_i\subset
\Omega$ acts as an affine transformation defined by real matrices $J_i$
of size $2$ and two components translation vectors ${\vec \kappa}_i$,
but the complementary subset $\Omega \setminus X_i$ is mapped out of
$\Omega$,  \begin{equation} f_i \left( \begin{array}{c} x\\ y
\end{array} \right) = \left\{ \begin{array}{c} J_i \left(
\begin{array}{c} x\\ y \end{array} \right) +{\vec \kappa_i} {\rm \quad
for \quad} x\in X_i  \\ {\rm \quad out \quad of \quad} \Omega {\rm \quad
for \quad} x\notin X_i \end{array} \right.  \label{IFS5} \end{equation}

Such functions $f_i$ are repelling, so instead of analyzing single
trajectories, generated by a random initial  point, it is convenient to
work with initially smooth probability measures.  Let us take for the
initial measure $\mu_0$ the uniform (Lebesgue) measure on $\Omega$.  To
iterate the repelling IFS 
we find the images of the measures covering each of the subsets $X_i$
by the maps $f_i$, $i=1,...,k$ and add all contributions with prescribed
weights -- probabilities $p_i$. More formally, we use the the dimensional
analogue of the Frobenius-Perron operator (\ref{FP1}) \begin{equation}
\rho'(x,y)=M[\rho(x,y)] = \sum_{i=1}^k p_i \bigl(f_i^{-
1}(x,y)\bigr)\rho\bigl(f_i^{-1}(x,y)\bigr)  \left|
\frac{\partial f_i(x,y)}{\partial (x,y)} \right|^{-1}  ,  \label{FP2}
\end{equation} where the contributions from the function $f_i$ have to
be taken into account only if the analyzed point $(x,y)$ belongs to
$f_i(X_i)$.  Then the preimage $f_i^{-1}(x,y)$ exists, so the $i$-th
term in the expression (\ref{FP2}) is meaningful.

\begin{figure}[hbt] \begin{center} 
\includegraphics[width=0.8\textwidth]{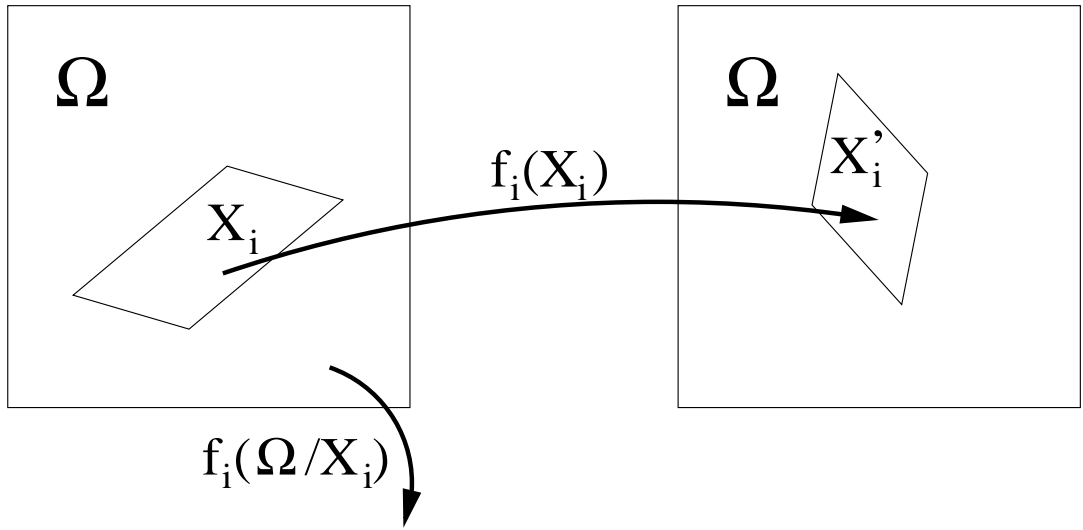} \end{center}
 \caption{Generalized IFS with 'repelling' functions $f_i$ which send
 the polygons $X_i$ into $ X_i'$, and the  remaining set
 $\Omega\setminus X_i$ is mapped out of the system.} \label{fig3}
 \end{figure}

As an example of such a generalized, repelling system we consider the
IFS ${\cal F}_5$ consisting of $k=13$ functions with equal
weights, $p_i=1/13$.  The detailed form of the affine functions
$f_i$ as well as the definitions of the sets $X_i$ are provided in the
Appendix. Figure 4 shows time evolution of the initially uniform measure
$\mu_0$.  The right hand side of the figure shows the measures
$\mu_n=M^n[\mu_0]$, while the left hand side shows the sets, where the
measure $\mu_n$ is concentrated. Initially uniform measure $\mu_0$
converges fast to the asymptotic measure $\mu_{*}$, and already after
$n=6$ iterations can hardly be distinguished from it.  The shape of the
support of $\mu_{*}$ explains why the name {\sl Euroattractor} has been
concocted for this system.

\begin{figure}[hbt] \begin{center}
\includegraphics[scale=0.99,height=360pt]{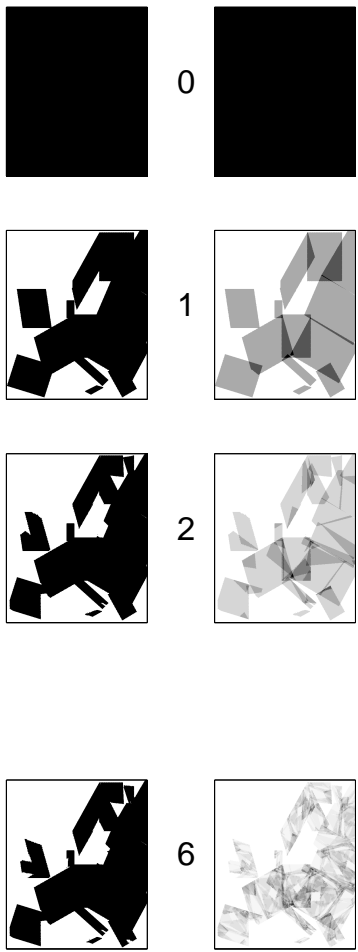}
\end{center} \caption{Euroattractor: the time evolution of the support
(left) and the measure (right) of the initially uniform measure $\mu_0$
in the square.  Already after $n=6$ iterations the measure cannot be
distinguished from the invariant measure $\mu_*$.} \label{fig4}
\end{figure}

\section{Closing Remarks}

The purpose of this work was to demonstrate usefulness of the concept of
{\sl Iterated Function System}.  This generalization of the concept of
a dynamical system incorporates  deterministic and stochastic behavior
and may be applied in various fields of physics, and in general, of
nonlinear science.  A possible extension of the concept of IFS for
quantum mechanics was recently worked out in \cite{LZS02}.

The IFS's belong to a larger class of random systems studied in
\cite{YOC91,PSV95}. From a mathematical point of view several
questions concerning the rigorous theory of IFS remain open. For
instance, the sufficient conditions, under which an IFS possesses a
unique, attracting invariant measure are known, but the formulation of
the necessary conditions remains still an exciting mathematical
challenge \cite{LY94,G92,GB95,E96}.  In this work we have introduced a
weakly contracting IFS's ${\cal F}_3$ and ${\cal F}_4$, which do not
satisfy the standard contractivity conditions, but do posses attracting
invariant measures. 

For any given set $S$ in the plane one may try to find an IFS, such that
its invariant measure $\mu_*$ has support on $S$.  This versatility of
IFS's allows one to apply them for  image synthesis, image processing
and image encoding \cite{Ba,BE88}.  For instance, the attractor of a
generalized IFS ${\cal F}_5$, described in this work, has the shape
resembling the contours of {\bf Europe}\footnote{Analyzing political
changes which took place in Europe during the last decade, one may ask,
whether European Union has enough attractive power to become a global
attractor for all countries in this continent.  The theory of nonlinear
systems prevents us from predicting the time evolution of an unstable
dynamical system in a long run.  Should we than just observe the events
waiting passively?  Or perhaps, taking the lesson from the Lorenz's
butterfly, could it be sufficient to flap the wings gently in the very
right moment?}.

It is a pleasure to thank Rados{\l}awa Bach and Wojciech
S{\l}omczy{\'n}ski  for constructive remarks and fruitful interaction.
This paper was presented at the  European Interdisciplinary School on
Nonlinear Dynamics {\sl Euroattractor 2002} organized in Warsaw in June
2002. We are grateful to W{\l}odzimierz Klonowski for an inspiring name
of the conference as well as for the opportunity to present this work
during that event. 

\appendix

\section{An Euroattractor}

In this appendix we provide an explicit definition of an {\sl
Euroattractor} -- a generalized, repelling IFS ${\cal F}_5$, the
invariant measure of which is shown in Fig.4.  The space $\Omega$ is
equal to an rectangle.  To address each pixel in a simple way its size is
taken to be $500 \times 600$.

The IFS consists of $k=13$ affine functions $f_i$ defined in
(\ref{IFS5}), and all the weights are constant, $p_i(x,y)=1/13$.
The parameters defining the corners of the quadrangles $X_i$, and the
elements of the transformation matrices $J_i$ and the translation
vectors ${\vec \kappa_i}$ are collected in Table 1.  Observe that some
entries of the Jacobi matrix $J_6$ have modulus larger than one, so this
IFS does not satisfy the contractivity condition. 

\newpage 

Table 1. Parameters defining an Euroattractor -- the IFS ${\cal F}_5$.

\begin{center}
\begin{tabular}{||c|c|c|c||} \hline \hline $i$ & $ X_i$ & $J_i$ & ${\vec
\kappa_i}$ \\ \hline \hline 1&$ \left( \begin{array}{c} 90\\ 370
\end{array} \right) \left( \begin{array}{c} 170\\   520 \end{array}
\right) \left( \begin{array}{c} 220\\   350 \end{array} \right) \left(
\begin{array}{c} 300\\   500 \end{array} \right) $&$ \left(
\begin{array}{cc} 0.37&0.66\\ -0.74&0.66 \end{array} \right) $&$ \left(
\begin{array}{c} -276\\ 371 \end{array} \right) $\\ \hline 2&$ \left(
\begin{array}{c} 520 \\  430 \end{array} \right) \left( \begin{array}{c}
340 \\  320 \end{array} \right) \left( \begin{array}{c} 420  \\ 530
\end{array} \right) \left( \begin{array}{c} 240  \\ 420 \end{array}
\right) $&$ \left( \begin{array}{cc}

-0.33&-0.73\\ 0.80&-0.59 \end{array} \right) $&$ \left( \begin{array}{c}
626\\ 355 \end{array} \right) $\\ \hline 3&$ \left( \begin{array}{c}
170\\   350 \end{array} \right) \left( \begin{array}{c} 265\\   480
\end{array} \right) \left( \begin{array}{c} 230\\   320 \end{array}
\right) \left( \begin{array}{c} 325\\   450 \end{array} \right) $&$
\left( \begin{array}{cc} 0.52&0.38\\ 0.01&0.68 \end{array} \right) $&$
\left( \begin{array}{c} 16\\ 218 \end{array} \right) $\\ \hline 4&$
\left( \begin{array}{c} 340\\   300 \end{array} \right) \left(
\begin{array}{c} 240\\   420 \end{array} \right) \left( \begin{array}{c}
500\\   400 \end{array} \right) \left( \begin{array}{c} 400\\   520
\end{array} \right) $&$ \left( \begin{array}{cc} -0.34&0.54\\
-0.61&-0.51 \end{array} \right) $&$ \left( \begin{array}{c} 192\\ 814
\end{array} \right) $\\ \hline 5&$ \left( \begin{array}{c} 380\\   500
\end{array} \right) \left( \begin{array}{c} 500\\   300 \end{array}
\right) \left( \begin{array}{c} 250\\   450 \end{array} \right) \left(
\begin{array}{c} 370\\   250 \end{array} \right) $&$ \left(
\begin{array}{cc} -0.17&-0.95\\ 0.96&0.08 \end{array} \right) $&$ \left(
\begin{array}{c} 812\\ 31 \end{array} \right) $\\ \hline 6&$ \left(
\begin{array}{c} 240\\   460 \end{array} \right) \left( \begin{array}{c}
340\\   550 \end{array} \right) \left( \begin{array}{c} 260\\   440
\end{array} \right) \left( \begin{array}{c} 360\\   530 \end{array}
\right) $&$ \left( \begin{array}{cc} 1.44&-1.05\\ -0.23&1.26 \end{array}
\right) $&$ \left( \begin{array}{c} 497\\ -24 \end{array} \right) $\\
\hline 7&$ \left( \begin{array}{c} 200 \\  490 \end{array} \right)
\left( \begin{array}{c} 50  \\ 590 \end{array} \right) \left(
\begin{array}{c} 150 \\  270 \end{array} \right) \left( \begin{array}{c}
0\\   370 \end{array} \right) $&$ \left( \begin{array}{cc} -0.31&-0.47\\
-0.98&0.22 \end{array} \right) $&$ \left( \begin{array}{c} 625\\ 97
\end{array} \right) $\\ \hline 8&$ \left( \begin{array}{c} 510\\   430
\end{array} \right) \left( \begin{array}{c} 460\\   540 \end{array}
\right) \left( \begin{array}{c} 305\\   330 \end{array} \right) \left(
\begin{array}{c} 245\\   440 \end{array} \right) $&$ \left(
\begin{array}{cc} -0.47&-0.032\\ 0.39&0.99 \end{array} \right) $&$
\left( \begin{array}{c} 485\\ -438 \end{array} \right) $\\ \hline 9&$
\left( \begin{array}{c} 230\\   280 \end{array} \right) \left(
\begin{array}{c} 230\\   450 \end{array} \right) \left( \begin{array}{c}
460\\   280 \end{array} \right) \left( \begin{array}{c} 460\\   450
\end{array} \right) $&$ \left( \begin{array}{cc} 0.26&1\\ -0.60&0.61
\end{array} \right) $&$ \left( \begin{array}{c} -20\\ 267 \end{array}
\right) $\\ \hline 10&$ \left( \begin{array}{c} 440\\   280 \end{array}
\right) \left( \begin{array}{c} 440\\   450 \end{array} \right) \left(
\begin{array}{c} 245\\   280 \end{array} \right) \left( \begin{array}{c}
245\\   450 \end{array} \right) $&$ \left( \begin{array}{cc}
-0.51&0.70\\ 0.87&0.41 \end{array} \right) $&$ \left( \begin{array}{c}
408\\ -339 \end{array} \right) $\\ \hline 11&$ \left( \begin{array}{c}
250\\   280 \end{array} \right) \left( \begin{array}{c} 310\\   190
\end{array} \right) \left( \begin{array}{c} 220\\   190 \end{array}
\right) \left( \begin{array}{c} 270\\   100 \end{array} \right) $&$
\left( \begin{array}{cc} -0.22& -0.37\\ 0.33& 0.11 \end{array} \right)
$&$ \left( \begin{array}{c} 439\\ 456 \end{array} \right) $\\ \hline
12&$ \left( \begin{array}{c} 330\\     0 \end{array} \right) \left(
\begin{array}{c} 200\\   190 \end{array} \right) \left( \begin{array}{c}
410\\   100 \end{array} \right) \left( \begin{array}{c} 280\\   290
\end{array} \right) $&$ \left( \begin{array}{cc} -0.60&-0.51\\
0.47&-0.38 \end{array} \right) $&$ \left( \begin{array}{c} 359\\ 187
\end{array} \right) $\\ \hline 13&$ \left( \begin{array}{c} 420\\    10
\end{array} \right) \left( \begin{array}{c} 420\\    90 \end{array}
\right) \left( \begin{array}{c} 330\\    10 \end{array} \right) \left(
\begin{array}{c} 330\\    90 \end{array} \right) $&$ \left(
\begin{array}{cc} 0&-0.31\\ 0.66&0 \end{array} \right) $&$ \left(
\begin{array}{c} 243\\ 30 \end{array} \right) $\\ \hline \hline
\end{tabular} \mbox{}
\end{center}


\end{document}